# An Adaptive-Parity Error-Resilient LZ'77 Compression Algorithm

Tomaz Korosec and Saso Tomazic, *Member, IEEE*

*Abstract*—The paper proposes an improved error-resilient Lempel-Ziv'77 (LZ'77) algorithm employing an adaptive amount of parity bits for error protection. It is a modified version of error resilient algorithm LZRS'77, proposed recently, which uses a constant amount of parity over all of the encoded blocks of data. The constant amount of parity is bounded by the lowest-redundancy part of the encoded string, whereas the adaptive parity more efficiently utilizes the available redundancy of the encoded string, and can be on average much higher. The proposed algorithm thus provides better error protection of encoded data.

The performance of both algorithms was measured. The comparison showed a noticeable improvement by use of adaptive parity. The proposed algorithm is capable of correcting up to a few times as many errors as the original algorithm, while the compression performance remains practically unchanged.

## I. INTRODUCTION

Lossless data compression algorithms, such as the Lempel-Ziv'77 (LZ'77) [1] algorithm and its variations, are nowadays quite common in different applications and compression schemes (GZIP, GIF, etc.). However, one of their major disadvantages is their lack of resistance to errors. In practice, even a single error can propagate and cause a large amount of errors in the decoding process. One possible solution for this problem is to use a channel coding scheme succeeding the source coding, which adds additional parity bits, allowing error correction and detection in the decoding process. However, such a solution is undesirable in bandwidth-limited systems, where the amount of bits required to carry some information should be as small as possible. A separate use of source and channel coding is not optimal, since it does not utilize inherent redundancy left by the source coding. This redundancy could be exploited for protection against errors. Therefore, joint source-channel coding seems to be a better solution. Several joint source-channel coding algorithms have been proposed in the past, e.g., [2], [3], and [4]. The redundancy left in LZ'77 and LZW encoded data and the possibility of using it to embed additional information has been considered and investigated in [5], [6], [7], and [8]. The LZRS'77 algorithm, proposed in [8], exploits the redundancy left by the LZ'77 encoder to embed parity bits of the Reed-Solomon (RS) code. Embedded parity bits allow detection and correction of errors with practically no degradation of the compression performance. However, due to the limited redundancy left in the encoded data, the ability to detect and correct errors is limited to a limited number of successfully corrected errors. To successfully correct $e$ errors, $2e$ parity bits should be embedded. In the above-mentioned scheme, the number of parity bits embedded in each encoded block is constant and equal for all blocks, thus e is limited by the redundancy of the block with the lowest redundancy.

In this paper, we propose an improvement to LZRS'77. Instead of keeping e constant, we change it adaptively in accordance with the redundancy present in the encoded blocks. In this way, we increase the average number of parity bits per block and thus also increase the total number of errors that can be successfully corrected. We named this new algorithm LZRSa'77.

The paper is organized as follows. In Section II, we briefly describe the LZRS'77 algorithm, which is the basis of the proposed adaptive-parity algorithm LZRSa'77 described in Section III.

T. Korosec and S. Tomazic are with the Faculty of Electrical Engineering, University of Ljubljana, Ljubljana, Slovenia, (email: tomaz.korosec@fe.uni-lj.si).

Experimental results comparing both algorithms are presented in Section IV. Some concluding remarks are given in Section V.

## II. Protection Against Errors Exploiting LZ'77 Redundancy

The basic principle of the LZ'77 algorithm is to replace sequences of symbols that occur repeatedly in the encoding string $\mathbf{X} = (X_1, X_2, X_3, \ldots)$ with pointers $\mathbf{Y} = (Y_1, Y_2, Y_3, \ldots)$ to previous occurrence of the same sequence. The algorithm looks in the sequence of past symbols $\mathbf{E} = (X_1, X_2, \ldots, X_{i-1})$ to find the longest match of the prefix $(X_i, X_{i+1}, \ldots, X_{i+l-1})$ of the currently encoding string $\mathbf{S} = (X_i, X_{i+1}, \ldots, X_N)$. The pointer is written as a triple $Y_k = (p_k, l_k, s_k)$, where $p_k$ is the position (i.e., starting index) of the longest match relative to the current index $i$, $l_k$ is the length of the longest match, and $s_k = X_{i+l}$ is the first non-matching symbol following the matching sequence. The symbol $s_k$ is needed to proceed in cases when there is no match for the current symbol. An example of encoding the sequence at position $i$ that matches the sequence at position $j$ is shown in Fig. 1.

To avoid overly large values of position and length parameters, the LZ'77 algorithm employs a principle called the sliding window. The algorithm looks for the longest matches only in data within the fixed-size window.

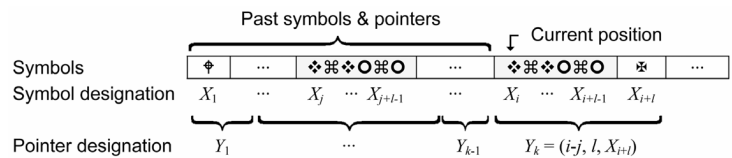

Fig. 1. An example of a pointer record for a repeated part of a string in the LZ'77 algorithm. The sequence of length $l = 6$ at position $j$ is repeated at position $i$, i.e., the current position.

Often, there is more than one longest match for a given sequence or phrase, which means more than one possible pointer. Usually, the algorithm chooses the latest pointer, i.e., the one with the smallest position value. However, selection of another pointer would not affect the decompression process. Actually, the multiplicity of matches represents some kind of redundancy and could be exploited for embedding additional information bits almost without degradation in the compression rate. A small decrease in compression performance could be noticed only in case when pointers are additionally Huffman encoded, as for example in GZIP algorithm, specified in [9]. With appropriate selection of one among $M$ possible pointers, we can encode up to $d = \lfloor \log_2 M \rfloor$ additional bits. These additional bits can be encoded with proper selection of pointers with multiplicity $M > 1$, as shown in Fig. 2. The algorithm LZS'77 that exploits the above-described principle in LZ'77 scheme was proposed and fully described in [5], [6], [7], and [8]. Since different pointer selection does not affect the decoding process, the proposed algorithm is completely backward compatible with the LZ'77 decoder.

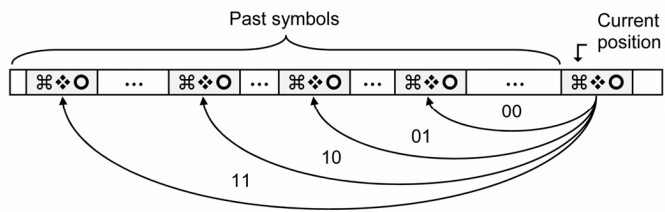

Fig. 2. An example of the longest match with multiplicity $M = 4$. With a choice of one of four possible pointers, we can encode two redundant bits.

The additional bits can be utilized to embed parity bits for error detection and correction. In [6] and [8], a new algorithm called LZRS'77 was proposed. It uses the additional bits in LZ'77 to embed parity bits of RS code originally proposed in [10]. In LZRS'77, an input string $\mathbf{X}$ is first encoded using the standard LZ'77 algorithm. Encoded data $\mathbf{Y}$ are then split into blocks of $255 - 2e$ bytes, which are processed in reverse order starting with the last block. When processing block $B_n$, $2e$ parity bits of block $B_{n+1}$ are computed first using RS(255, $255 - 2e$) code and then those bits are embedded in the pointers of block $B_n$ using the previously mentioned LZS'77 scheme. Parity bits of the first block can be stored at the beginning of the file if we also wish to protect the first block. Otherwise, to assure backward compatibility with the LZ'77 decoder, protection of the first block should be omitted.

In the decoding process, the procedure is performed in the opposite order. The first block is corrected (only in the case when the first block is protected as well) using parity bits appended at the beginning of the file. Then it is decompressed using the LZS'77 decompression algorithm, which reconstructs the first part of the original string and also recovers parity bits of the second block. The algorithm then corrects and decompresses the second block and continues in this manner till the end of the file.

The desired maximum number of errors $e$ to be effectively corrected in each block during the decoding process is given as an input parameter of the algorithm. This number is upward-limited by the ability to embed bits in the pointer selection, i.e., by the redundancy of the encoded data. In the LZRS'77 algorithm, $e$ is constant over all blocks; thus its value is limited by the block with the lowest redundancy.

### III. The LZRSa'77 Algorithm with Adaptive Parity

A constant $e$ over all encoding blocks, as in LZRS'77, is not optimal, since redundancy in different parts of data string can differ significantly. If there is just one part of the string that has very low redundancy, it will dictate the maximum value of $e$ for the whole string. Such low-redundancy blocks are usually at the beginning of the encoded data, since there are not yet many previous matches that would contribute to redundancy. Better utilization of overall redundancy would be possible with an adaptive $e$, changing from one block to another according to availability of redundancy bits in each block. In that case, low-redundancy parts of the string would affect the error protection performance just of these parts, whereas the rest of the string could be better protected according to its redundancy availability. As a result, the value of $e$ is still upward-limited by the overall redundancy but its average value can be higher, resulting in better resistance to errors.

On the basis of the above-described assumptions, we propose an improved version of the LZRS'77 algorithm, named LZRSa'77, where 'a' refers to adaptive $e$. The input string $\mathbf{X}$ is first encoded using the standard LZ'77 algorithm, when the multiplicity $M_k$ of each pointer is also recorded. The encoded data is then divided into blocks of different lengths, according to the locally available redundancy. Firstly, $255–2e_1$ bytes are put in the first block $B_1$, where $e_1$ is given as an input parameter of the algorithm. Then, the number of parity bits $2e_2$ of the second block $B_2$ is calculated, where $e_2$ is given as:

$$e_2 = \left\lfloor \sum_{k \in B_1} \left\lfloor \log_2 M_k \right\rfloor / 16 \right\rfloor. \tag{1}$$

If, for example, the number of additional bits that could be embedded in the pointers multiplicity of the first block ($\sum \log_2 M_i$) is 43, then the number of parity bits of the second block would be $2e_2 = 2\lfloor 43/16 \rfloor = 4$. According to the obtained value, the second block length is $255–2e_2 = 251$ bytes. The process is then repeated until the end of the input data is reached. We obtain $b$ blocks of different lengths $255–2e_n$.

After dividing all the data into blocks of different lengths, the process of RS coding and embedding of parity bits is performed. The blocks are processed in reverse order, from the very last to the first, as with the LZRS'77 algorithm. The number of parity bits $2e_n$ for RS coding varies for each block. The sequence of operations of the encoder is illustrated in Fig. 3.

As mentioned above, the desired error correction capability of the first block $e_1$ is given as an input parameter of an algorithm, whereas $e_n$ for all the other blocks are obtained from the redundancy of their preceding blocks and are as high as the redundancy permits. As in the LZRS'77 algorithm, parity bits of the first block are appended at the beginning of the encoded data, or omitted if we want to preserve backward compatibility with the standard LZ'77 decoder. In the last case, $e_1$ is equal to zero.

The decoding process is similar to that used in the LZRS'77 decoding algorithm. Each block $B_n$ is first error-corrected using $2e_n$ parity bits known from the previous block $B_{n-1}$, then decoded using the LZS'77 decoder to decompress some of the original string and obtain $2e_{n+1}$ parity bits of the next block. The amount of parity bits is used to determine the length of the next block $B_{n+1}$, whereas the parity bits themselves are used to correct the block. The process is continued to the last block. A high-level description of the encoding and decoding algorithms is shown in Fig. 4.

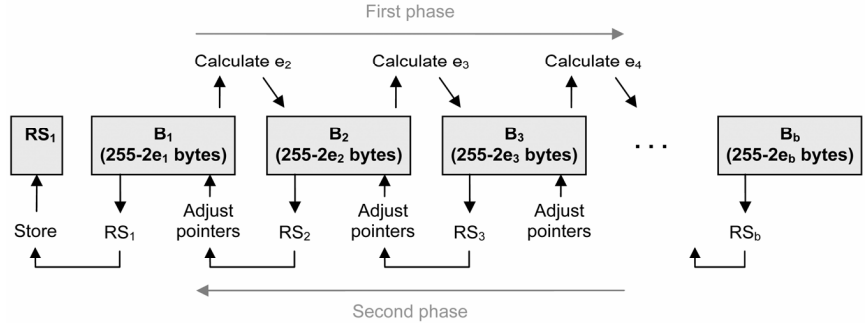

Fig. 3. The sequence of operations on the compressed data as processed by the LZRSa'77 encoder. Here $RS_n$ are parity bits of the block $B_n$.

## IV. EXPERIMENTAL RESULTS

To evaluate the performance of the proposed algorithm, we performed several tests with different files from the Calgary corpus [11]. We implemented our proposed algorithm in the Matlab 6.5.1 Release 13 program tool. For the basic LZ'77 encoding, the LZ'77 algorithm with a sliding-window length of 32 kilobytes was used. It was implemented in Matlab as well. Maximum length of pointers was chosen to be 255 bytes.

LZRSA'77_ENCODER $(X, e_1)$

**let** $b, j \leftarrow 1, 0$
$[P, p] \leftarrow$ LZ'77_COMPRESS$(X)$
**while** $j < |P|$ **do**
        append $(j + 1)...(j + 255 - 2e_b)$ bytes of $P$ to $B_b$
        **let** $j \leftarrow j + 255 - 2e_b$
        **let** $b \leftarrow b + 1$
        **evaluate** $e_b$ by counting possible pointers in $p$ for
                $B_{b-1}$
**for** $n \leftarrow b,...,2$ **do**
        **let** $RS_n \leftarrow$ RS_ENCODER$(B_n, e_n)$
        **embed** in the block $B_{n-1}$ the bits $RS_n$ using LZS'77
**let** $RS_1 \leftarrow$ RS_ENCODER$(B_1, e_1)$
**let** $B \leftarrow (B_1, B_2,..., B_b)$
**return** $e_1, RS_1, B$

LZRSA'77_DECODER $(e_1, RS_1, B)$

$D \leftarrow$ empty string
**let** $B_1 \leftarrow$ first $255 - 2e_1$ bytes of $B$
**let** $j \leftarrow 255 - 2e_1 + 1$
**let** $n \leftarrow 2$
**if** RS_DECODER$(B_1 + RS_1, e_1)$ = errors
        **then** correct $B_1$
**append** LZ'77_DECOMPRESS$(B_1)$ to $D$
**while** $j < |B|$ **do**
        **recover** $RS_n$ from the pointers in $B_{n-1}$ using LZS'77
        **let** $e_n \leftarrow$ half a number of $RS_n$ bytes
        **let** $B_n \leftarrow$ next $255 - 2e_n$ bytes of $B$ from index $j$ on
        **let** $j \leftarrow j + 255 - 2e_n$
        **let** $n \leftarrow n + 1$
        **if** RS_DECODER$(B_n + RS_n, e_n)$ = errors
                **then** correct $B_n$
        **append** LZ'77_DECOMPRESS$(B_n)$ to $D$
**return** $D$

Fig. 4. The error-resilient LZ'77 algorithm with adaptive parity $2e_n$. Here $X$ is the input string, $e_1$ is the maximum number of errors that can be corrected in the first block, $P$ is the LZ'77 encoded string of pointers, $p$ is a vector of possible positions for each pointer, $B_n$ are blocks of encoded data of variable length $255-2e_n$, $RS_n$ are RS parity bits of the block $B_n$, and $D$ is the recovered string.

In the experiment, we first compared the maximal value of constant $e$ and average value of an adaptive $e$ ($E(e_n)$) in different test strings. For this purpose, we encoded different files from the Calgary corpus using the LZRS'77 and LZRSa'77 algorithms. For maximal $e$ observation, we performed tests only on strings of 10,000 bytes length, since the lowest-redundancy parts proved to be in the first blocks of the encoded strings, because there are not so many past symbols. Thus, different string lengths do not affect the maximal $e$, as long as the beginning of the string is the same. For this reason, we rather performed tests for different substrings of the same length within each file starting at different positions. Average maximal $e$ obtained over all tested substrings for each file is given in the second column of Table I, whereas maximal $e$ of the first substring of each file (and thus that corresponding to the whole file) is given in the third column. Even if, in an unexpected case, the lowest redundancy part of the whole file is not within the first 10,000 symbols, the obtained results were still relevant, since we made additional experiments on error-correction performance on the first 3000 and 30,000 symbols with the same constant parity used.

When observing $E(e_n)$, we performed measurements on two different lengths of source strings, i.e., 3000 bytes and 30,000 bytes, and we again performed the tests on different substrings within each file for both lengths. The value of $e_1$ was in all cases equal to 1. Results are shown in fourth and fifth columns of Table I.

The experiment results showed that the maximal constant $e$ that could be embedded in the redundancy of the encoded string is in the best case equal to 3, whereas average adaptive $e$ over large number of blocks could be from 4.5 up to 8. These results already justify the use of adaptive $e$. To justify it further, we performed another experiment. We tested the ability of each algorithm to correct random errors.

TABLE I

VALUES OF MAXIMAL CONSTANT AND AVERAGE ADAPTIVE $e$ FOR DIFFERENT LENGTH ($L$) SUBSTRINGS OF THE CALGARY CORPUS FILES

| file name | $E(e_{max})$ over substrings with $L$=10,000 | $e_{max}$ of the whole file | $E[E(e_s)]$ over substrings with $L$=3000 | $E[E(e_s)]$ over substrings with $L$=30,000 |
|---|---|---|---|---|
| `bib` | 2.00 | 2 | 4.79 | 5.29 |
| `book1` | 2.38 | 2 | 4.75 | 4.94 |
| `book2` | 2.18 | 1 | 4.64 | 5.04 |
| `geo` | 2.40 | 3 | 5.48 | 8.32 |
| `news` | 1.92 | 1 | 5.05 | 5.93 |
| `obj1` | 2.50 | 2 | 5.05 | / |
| `obj2` | 1.46 | 1 | 4.68 | 6.77 |
| `paper1` | 2.00 | 1 | 4.64 | 5.14 |
| `paper2` | 1.88 | 1 | 4.65 | 4.80 |
| `paper3` | 1.75 | 1 | 4.62 | 4.87 |
| `paper4` | 1.00 | 1 | 4.70 | / |
| `paper5` | 1.00 | 1 | 4.75 | / |
| `paper6` | 1.67 | 1 | 4.81 | 5.14 |
| `progc` | 2.00 | 2 | 4.65 | 5.70 |
| `progl` | 2.00 | 2 | 4.48 | 6.21 |
| `progp` | 2.25 | 2 | 4.96 | 5.69 |
| `trans` | 1.22 | 2 | 4.82 | 6.26 |

When testing error correction performance, we performed measurements on three different files from Calgary corpus, i.e., `news`, `progp`, and `geo`, which allow maximal values of constant $e$ equal to 1, 2, and 3 respectively, as shown in Table I. Measurements were made on the first 3000 and 30,000 bytes of each file respectively. When using the LZRSa'77 algorithm, $e_1$ could be an arbitrary value. However, we chose values that approximately correspond to $E(e_n)$ for each of the tested files. Thus, we chose $e_1 = 5$ for the `news` and `progp` test strings, and $e_1 = 8$ for the `geo` test string.

We tested the resilience to errors by introducing different number of errors randomly distributed over the whole encoded string. For error generation, we used a built-in Matlab function, called `randerr`, which generates patterns of geometrically distributed bit errors.

Results for the three test strings, all in two different length variations, and for both described algorithms used (LZRS'77 and LZRSa'77) are shown in the graphs in Fig. 5 to Fig. 7. Each case of string type, string length and algorithm used was tested with different numbers of injected errors.

For each number of errors, 100 trials with different randomly distributed errors were performed and number of successful data recovers tested.

In the graphs in Fig. 5 to Fig. 7, the measured results are plotted with discrete points, whereas continuous curves represent a polynomial-fitted approximation. The results show quite an improvement in error correction capability when using the LZRSa'77 algorithm instead of LZRS'77, which is a direct consequence of the larger amount of parity used in the first algorithm. The performance improvement decreases with increasing constant $e$ from 1 to 3, but is still noticeable also in the last case, which is practically the best we could achieve with the LZRS'77 algorithm. As can be seen from the results, the performance improvement also somewhat increases with increasing length of the string. This is probably due to the increasing $E(e_n)$ with increasing length of the string, as evident from Table I, whereas constant $e$ remains the same.

The performance of the LZRSa'77 algorithm could be slightly further improved using higher value of $e_1$, which would, however, improve only the protection of the first block.

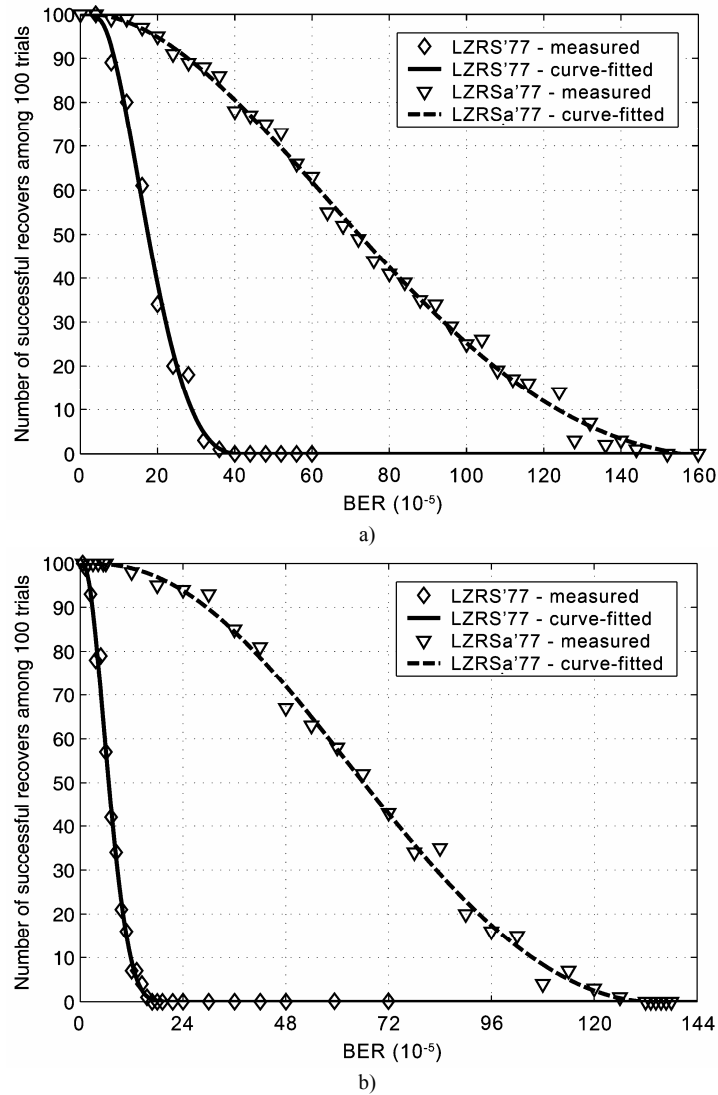

Fig. 5. The number of successful recovers among 100 trials for two different length ($L$) substrings of the file news, for increasing number of bit errors geometrically distributed over the encoded strings, represented as Bit Error Rate (BER), end different algorithm used (LZRS'77 and LZRSa'77). a) $L = 3000$ bits; b) $L = 30{,}000$ bits.

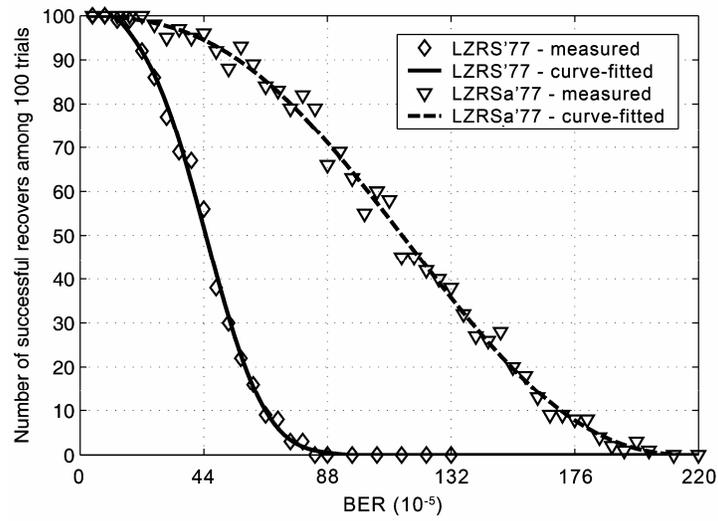

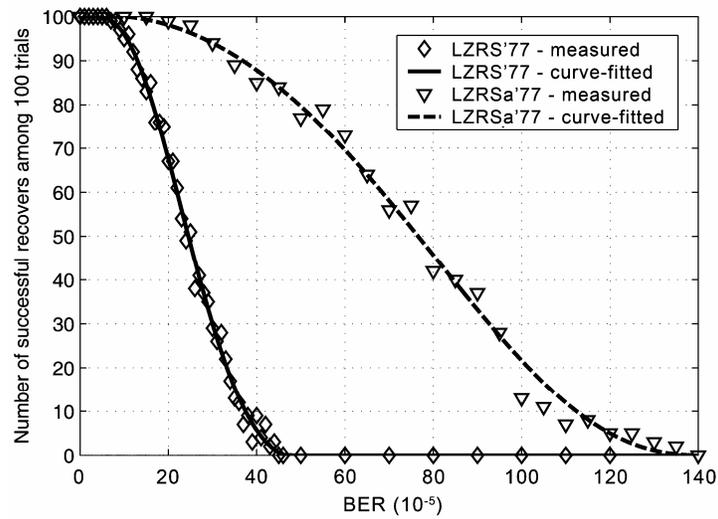

Fig. 6. The number of successful recovers among 100 trials for two different length (*L*) substrings of the file `progp`, for increasing number of bit errors geometrically distributed over the encoded strings, represented as BER, end different algorithm used (LZRS'77 and LZRSa'77). a) *L* = 3000 bits; b) *L* = 30,000 bits.

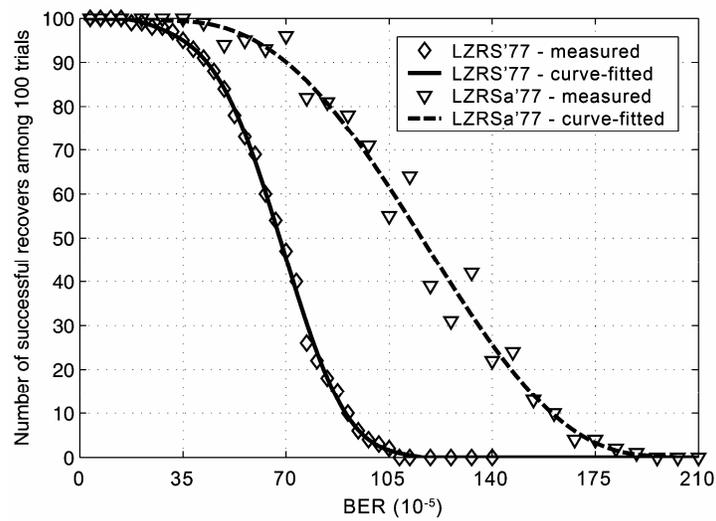

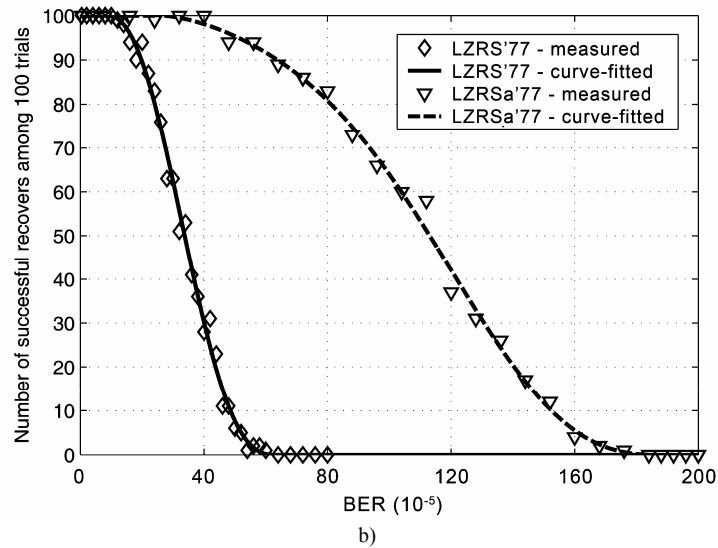

b)

Fig. 7. The number of successful recovers among 100 trials for two different length ($L$) substrings of the file `geo`, for increasing number of bit errors geometrically distributed over the encoded strings, represented as BER, end different algorithm used (LZRS'77 and LZRSa'77). a) $L = 3000$ bits; b) $L = 30,000$ bits.

## V. Conclusion

An improved version of the error-resilient LZ'77 data compression scheme was presented. It allows use of adaptive number of parity bits over different blocks of encoded data according to available redundancy in the blocks. Compared to the recently proposed LZRS'77 scheme allowing only constant number of parity bits along the whole string, the new solution better utilizes available redundancy in the string, resulting in a larger number of errors that can be effectively corrected. Such an improvement does not practically degrade the compression rate compared to the LZRS'77 algorithm. Even though the parity of each block has to be calculated each time from the redundancy of the previous block, the time complexity of the new algorithm remains on the order of that of the LZRS'77 algorithm.

However, some legacy from the LZRS'77 algorithm still remains in the new algorithm and represents two unsolved problems. The first is a question of an online encoding process, which could not be achieved due to the reverse order of block processing. The second is protection of the first block while maintaining backward compatibility.